# Competing Interactions and Magnetic Frustration in $Yb_4LiGe_4$

S. M. Disseler[1], J. N. Svensson[1*], S. C. Peter[2], C. P. Byers[3†], C. Baines[4], A. Amato[4], S. R. Giblin[5], P. Carretta[6], and M. J. Graf[1‡]

[1] Department of Physics, Boston College, Chestnut Hill, MA 02467 USA

[2] New Chemistry Unit, Jawaharlal Nehru Centre for Advanced Scientific Research, Jakkur, Bangalore-560 064, India

[3] Department of Physics, Marietta College, Marietta, OH 45750 USA

[4] Paul Scherrer Institute, CH 5232 Villigen PSI, Switzerland

[5] Rutherford Appleton Laboratory, Didcot, Oxfordshire OX11 0QX, UK

[6] Department of Physics "A. Volta" and CNISM, University of Pavia, I27100 Pavia, Italy

**ABSTRACT**

We have studied polycrystalline $Yb_4LiGe_4$, a ternary variant of the $R_5T_4$ family of layered compounds characterized by a very strong coupling between the magnetic and crystallographic degrees of freedom. The system is mixed valent, with non-magnetic $Yb^{2+}$ and magnetic $Yb^{3+}$ present, and is characterized by coexisting ferromagnetic and antiferromagnetic correlations. We present measurements of resistivity, AC-susceptibility, specific heat, and muon spin relaxation (μSR), below 1 K. The low temperature measurements suggest a transition to a mesoscopically inhomogeneous magnetically ordered state below 2 K characterized by fluctuations well below the ordering temperature. This unusual state is believed to result from the enhanced two-dimensionality produced by Li substitution and frustration effects inherent in the Yb sub-lattice geometry.

# I. INTRODUCTION

Much of current condensed matter physics centers on the unusual emergent behaviors that occur in complex systems with competing interactions. The prototypical example would be heavy fermion systems [1], where the tendency towards magnetic order facilitated by the RKKY interaction competes with Kondo screening of local moments by conduction electrons. Emergent properties in the heavy fermion systems include greatly enhanced carrier effective masses on the order of hundreds of electron masses and unconventional superconductivity. Similarly, magnetic systems with strong interactions promoting magnetic order but with a lattice structure or competing interactions that frustrate that order can exhibit highly degenerate ground states with unusual properties [2]. Because of the strong competition between interactions, both heavy fermion systems and magnetically frustrated systems have ground states that can be tuned by slight modifications of the internal or external parameters, e.g., lattice deformations, chemical substitution, or the application of a magnetic field. For example, a family of ternary compounds was observed to support examples of both heavy fermion and frustrated behavior [3]. Theoretical works are also underway to describe the possible ground states when geometric frustration occurs within a Kondo lattice system, with emphasis on Yb-based systems [4]. It is of great interest to expand the range of systems with competing interactions to investigate the possibility of new emergent behaviors.

The $R_5T_4$ family of compounds ($R$ = rare earth, $T$ = Ge or Si) are well-studied in the magnetism community, primarily for the range of magnetic ordering displayed near room temperature, and for an unusually large magnetothermal effect which has potential for refrigeration applications [5]. The materials are layered, and the crystal structures are variations of the orthorhombic $Sm_5Ge_4$ type. The wide range of observed magnetic states is believed to result from a high sensitivity of the magnetic interaction to slight variations in the bonding states, and in particular the formation of $T$-$T$ dimers connecting the layers. Moreover, we note that the $R$ sites form a Shastry-Sutherland lattice in two dimensions in the crystalline a-b plane (see crystal structure in Fig. 1), which is a configuration that frustrates magnetic order [3, 4, 6]. These competing interactions play an important role in $Gd_5Ge_4$, for example. This material has competing intralayer ferromagnetic correlations and interlayer antiferromagnetic correlations, which results in complex magnetic behavior. The system undergoes a Griffiths phase transition ($T_G$ = 240 K) into ferromagnetic clusters [7], and long-range antiferromagnetic order is observed for temperatures below the Néel temperature $T_N$ = 128 K. At lower temperatures a Martensitic-like structural transition facilitates a transition from AFM to FM order. While most of the $R_5T_4$ materials exhibit either ferro- or antiferromagnetic order at temperatures ranging from tens to hundreds of degrees Kelvin, one member of the family, $Yb_5Ge_4$, has a reported onset of antiferromagnetic order at roughly 2 K which has only been observed via Mossbauer spectroscopy [8]. The system is mixed-valent, with the Yb assuming the non-magnetic $Yb^{2+}$ and magnetic $Yb^{3+}$ valence states in comparable proportions. The existence of several competing interactions with comparable magnitude, as evidenced by the low transition temperature, is expected to result in unusual properties.

In this work we study a derivative material, $Yb_4LiGe_4$. Detailed studies of the chemistry and structure of this material [9, 10] have shown that the Li substitutes for the Yb in lattice sites such that the layers containing $Yb^{3+}$ ions are separated by Li containing layers (see Fig. 1), suggesting that the material will tend towards two-dimensional behavior. Measurements of specific heat, $^7Li$ spin-lattice relaxation, and magnetic susceptibility demonstrated that the system was characterized by coexisting ferromagnetic and antiferromagnetic correlations [10]. In this work, we extend our studies to lower temperatures, and present measurements of resistivity, AC susceptibility, specific heat, and muon spin relaxation (μSR) to below $T$ = 1 K. We find evidence for the onset of unusual magnetic ordering below 2 K. The persistence of magnetic fluctuations to 100 mK as determined by the μSR data imply that quantum fluctuations play an important role in this system.

## II. SAMPLE PREPARATION AND CHARACTERIZATION

A detailed description of our fabrication and characterization of $Yb_4LiGe_4$ and $Yb_5Ge_4$ samples is presented elsewhere [10], and is generally consistent with the results reported by Xie *et al.* [9]. We present an overview of the process here. The starting materials were ingots of ytterbium (99.99 wt%), lithium rods (99.4 wt%), and germanium pieces (99.9999 wt%). The elements in stoichiometric ratios were heated in closed tantalum tubes to roughly 1070 K using a high frequency (HF) furnace. Careful handling of the ytterbium and lithium was necessary in order to minimize impurities, and was carried out inside an argon glove-box system. The synthesized samples were loaded into a planetary ball mill for milling of about half an hour. Phase analysis was done by powder x-ray diffraction and Scanning Electron Microscopy (SEM), coupled with Energy Dispersive X-ray Spectroscopy (EDS). $Yb_4LiGe_4$ and $Yb_5Ge_4$ samples after HF treatment are polycrystalline in nature, light grey with metallic luster and air sensitive. The powder x-ray pattern reveals the presence of the secondary phases $Yb_4Ge_3$, Yb and Ge in $Yb_4LiGe_4$ and $Yb_5Ge_3$ and Ge in $Yb_5Ge_4$. In order to avoid these impurities, the products were initially treated in a ball miller for homogeneity. The powder XRD patterns of the samples obtained before and after milling were identical. Powder blends obtained from the milling process were then loaded again into Ta tubes and heated up to 1070 K using HF furnace. After the second step of heat treatment the resulting products were found to contain less than 1% impurity phase based on the resolution of the diffractometer. The detailed analysis of the x-ray patterns [9, 10] show that the crystal structure is slightly distorted from the $Yb_5Ge_4$ structure, and, importantly, the Li substitutes for Yb at the Yb3 position as shown in Fig. 1, resulting in a layer with no magnetic ions and potentially reducing the dimensionality of the system.

In order to make uniform pellets, the resulting pieces of $Yb_4LiGe_4$ were ball milled once again, and the powder was loaded into a graphite die lined with Ta foil in a high purity argon atmosphere. The powder was pressurized to 140 MPa, and the temperature ramped to 700 $^{o}$C over a 5 minute period and held at that temperature for an additional 5 minutes before releasing the pressure and cooling the sample. These pellets were once again confirmed to be single phase by x-ray diffraction. This heat press technique was found to introduce impurities into the $Yb_5Ge_4$ material, so susceptibility measurements on that material were made on pieces produced by the technique described in the previous paragraph, while µSR measurements were made on pieces that were powdered once again via ball milling and confirmed to be single phase.

To confirm the consistency of our sample quality, AC susceptibility and resistivity data below 2 K were taken for the original sample batch (for which data was presented in Ref. 10) and found to be consistent with the data presented here. Moreover, above 2 K we have some overlapping resistivity, specific heat, magnetization, and magnetic susceptibility data with data for our original sample batch, and these are also consistent.

.

## III. RESULTS

**A. Resistivity**

Electrical resistivity of several samples was measured over the range 100 mK < $T$ < 300 K using the standard 4-point method utilizing both AC and DC techniques. The current excitation level was varied to ensure that there was no self-heating of the sample at low temperatures. At room temperature $Yb_4LiGe_4$ is a conductor, consistent with electron localization function calculations [9], although it exhibits a very large room temperature resistivity of 1500 μΩ-cm, much larger than most Yb intermetallic compounds. As shown in the inset of Fig. 2 [10], below room temperature the resistivity increases rapidly, reaching a broad maximum at 100 K; further cooling reveals a weakly decreasing resistivity down to low temperatures. We note that such a maximum in the resistivity is observed in many Kondo-lattice systems. Below 2 K, we observe broad maximum that exhibits negative magnetoresistance and is suppressed by a field of 2 T. As shown below, we associate this lower temperature maximum with the onset of magnetic order.

**B. Magnetization and Magnetic Susceptibility**

The DC magnetization was measured over temperature range 2 K < $T$ < 10 K using the commercial Oxford MagLab Measurement System and Quantum Design PPMS. This was calibrated with a superconducting Nb sphere. AC susceptibility measurements below 3 K were carried out using an astatic three-coil setup constructed for use in a $^3$He refrigerator. The drive frequency was varied between 150 Hz and 15 kHz, and typical oscillating field amplitudes were on the order 0.1 Oe, and confirmed to not induce heating in the sample.

Our earlier susceptibility measurements [10] on $Yb_4LiGe_4$ demonstrated paramagnetic Curie-Weiss behavior for $T$ > 60 K, with a small negative $\theta_{CW}$ = -3.0(1) K, indicative of weak antiferromagnetic correlations. This value is consistent with the reported value of $\theta_{CW}$ = -4(2) K for $Yb_5Ge_4$ [11]. The slope yielded an effective magnetic moment $\mu_{eff}$ = 7.45(10) $\mu_B$ per formula unit, or 3.7 $\mu_B$ per Yb, significantly larger than the 2.73 $\mu_B$ per Yb reported for $Yb_5Ge_4$ and 2.84 $\mu_B$ per Yb for $Yb_5Si_4$ as reported in Ref. 11, and the 2.7 $\mu_B$ per Yb as reported in Ref. 12 for related compound $Yb_4MgGe_4$. The estimated fraction of magnetic $Yb^{3+}$ sites was extracted from the effective moment following the procedure in Ref. 11, and found to be $n_{3+}$ = 0.67, close to the 0.57 value extracted from x-ray absorption spectroscopy (XAS) measurements at room temperature [10], and roughly 30% greater than the fraction reported for $Yb_5Ge_4$ [8]. Tobash and Bobev [12] have described the isostructural compound $Yb_4MgGe_4$ as alternating layers of the virtual compounds $Yb_2MgGe_2$ and $Yb_2Ge_2$ along the b-direction, with a charge arrangement approximated by $(Yb^{2+})_2Mg^{2+}(Ge_2^{6-})$ and $(Yb^{3+})_2(Ge_2^{6-})$ following the Zintl Klemm concept [9, 12]. A consequence of substituting Li for Mg in this structure is that to maintain adequate charge balance, there should be an increase in the average valence of the Yb in the $Yb_2LiGe_2$ psuedo-plane, or a slight undercharging of the Ge dimers leading to bond instabilities. The results from susceptibility appear to confirm the former, as we observe an increased $\mu_{eff}$ and $n_{3+}$ for $Yb_4LiGe_4$.

Isothermal magnetization measurements on $Yb_4LiGe_4$ taken in the range 2 K < $T$ < 20 K, as shown in Fig. 3a, show no sign of saturation of the Yb moments, and a yield a maximum extrapolated high field moment of 1.1 $\mu_B$/Yb atom. This is smaller than the value determined from our Curie-Weiss extrapolation, which could be due to the presence of strong correlations [13], although extracting any firm conclusion is difficult because our samples are polycrystalline while the magnetic properties could show significant crystalline anisotropy

AC-susceptibility measurements below $T$ = 2 K in zero applied field exhibit strong deviations from the Curie-Weiss behavior observed at higher temperature. Indeed, we observe a peak at $T_{susc}$ = 1.45 K, as shown in Fig. 4. The peak observed for $Yb_4LiGe_4$ shows a dramatic reduction in height when small magnetic fields (of tens of gauss) are applied, which is not expected for any system undergoing traditional disorder-order transition. This peak showed no hysteresis effects, and was independent of cooling history (that is, field cooled versus zero field cooled). Variation of the drive frequency between 150 Hz and 15 kHz resulted in an increase $T_{susc}$ by approximately 1 %. As the samples examined were all bulk polycrystals, we cannot comment on the anisotropy of the susceptibility; however as nearly all of the $R_5T_4$ compounds show highly anisotropic behavior, it would be reasonable to assume that the effect of the magnetic field is to suppress ordering along some preferred axes. Measurements of single crystal samples are required to further investigate this behavior.

In contrast, reported DC susceptibility measurements for $Yb_5Ge_4$ exhibit a broad peak at a temperature of 3.2 K, nearly twice that of the ordering temperature (as determined from Mossbauer spectroscopy) [8]; susceptibility data were not taken through the ordering temperature in that work. For comparison, then, we measured the susceptibility of $Yb_5Ge_4$ over the temperature range 0.5 K < $T$ < 5 K for a sample fabricated by us, and observe a broad peak near $T$ = 2.9 K (see inset, Fig. 4), very similar to that reported in Ref. 11. On the other hand, although the susceptibility decreases rapidly with decreasing temperature below this maximum, we find no sharp features that would signal the onset of magnetic order at lower temperatures. This result is inconsistent with the Mossbauer measurements showing the abrupt onset of magnetic order at 1.7 K.

## C. Specific Heat

Previously we had measured the specific heat for $Yb_4LiGe_4$ in the temperature range 3 K $< T < 60$ K [10], and found results similar to those for $Yb_5Ge_4$ [11]. For both materials, an increase in the specific heat below 10 K is attributed to magnetic degrees of freedom, as no such increase is observed in the nonmagnetic isostructural compound $Lu_5Ge_4$ [11]. Applied magnetic fields to 3 T were observed to have little effect on $Yb_4LiGe_4$ in this region.

The low temperature 0.7 K $< T < 3.0$ K specific heat was measured in a semi-adiabatic calorimeter built for use in our $^3$He refrigerator; this data matches smoothly with the data taken on a different sample above 3 K and presented in Ref. 10. The addenda were measured separately, and confirmed to be less than 2 % of the total heat capacity in this temperature range. Data were taken in zero field, 100 G, and 500 G, and are shown in Fig. 5. We find a sharp peak at $T_{spht} = 1.75$ K that clearly signals a phase transition. A relatively rapid increase of specific heat with decreasing temperature is observed near 2.2 K. The nature of this increase is unclear at present, although other $R_5T_4$ samples exhibit structural changes before undergoing a magnetic transition [7], and so the possibility of a structural transition warrants further study. However, we note that $Yb_2O_3$ has a magnetic transition at 2.3 K [14], and we cannot rule out trace amounts of that impurity material causing a small anomaly superimposed on the rising $Yb_4LiGe_4$ specific heat. In contrast to the transition observed in AC susceptibility, the magnitude of the transition at 1.75 K is essentially unaffected by applied magnetic fields up to 500 G. Also to be noted is the large value of $C/T$ for $T > 1.75$ K, which is nearly 0.5 $J/K^2$mole-Yb, an extremely large value characteristic of heavy fermion systems. Fitting C/T below 1.5K with the standard $\gamma + \beta T^2$, and extrapolating to zero temperature, we estimate that the total change in entropy between 0 K and 1.75 K is approximately 0.6 J/Kmole-Yb, or 10% Rln2. Since 67% of the Yb ions are magnetic, a more informative result is 15 % $R$ln2 per mole $Yb^{3+}$. While this large value links the peak in the specific heat to changes in the magnetic degrees of freedom, it is clear that we have only achieved partial ordering, with significant entropy associated with the low temperature state.

### D. Muon Spin Relaxation (μSR)

In order to probe the local magnetic field to very low temperatures and to execute a comparative study between the $Yb_5Ge_4$ and $Yb_4LiGe_4$ we have performed μSR measurements. In μSR polarized positive muons are implanted in the sample, and the local magnetic field at the muon stopping site(s) is probed via the averaged normalized time-dependent response (depolarization) of the muon spin orientation; the measurement is sensitive to both quasi-static and fluctuating magnetic fields. A description of the technique is given in References 15 – 17. The μSR measurements were performed on the MUSR spectrometer at the ISIS pulsed beam facility at the Rutherford Appleton Laboratories, and the GPS and LTF spectrometers on the πM3 continuous beamline at Paul Scherrer Institute (PSI). The same $Yb_4LiGe_4$ samples were measured at both facilities and similarly mounted on silver plates for use in dilution refrigeration systems to 20 mK, and at ISIS in a gas flow cryostat. Samples of $Yb_5Ge_4$ were powdered and sealed inside metalized Mylar film, and measured in the GPS spectrometer only in a gas flow cryostat. Zero field (ZF) and longitudinal field (LF) measurements were performed at both facilities up to 500 G, while limited 100 G transverse field (TF) measurements were performed at PSI.

Data for $Yb_4LiGe_4$ in ZF and 500 G LF are shown for several temperatures in Fig. 6. At high temperatures $T > 10$ K the muon depolarization for both samples is well-described by a single exponential decay, which we associate with fluctuating magnetic field of the $Yb^{3+}$ ions, with a possible small quasi-static contribution from the $^7Li$ nuclei for $Yb_4LiGe_4$. At very low temperatures ($T = 20$ mK) and zero field we found no spontaneous oscillations for the data taken at PSI (see inset Fig. 7), indicating that any magnetic order, if present, is either characterized by a small local moment, or is incommensurate with the lattice. For the ZF depolarization curves it was not possible to fit the low temperature data with a single component depolarization function. This is supported by TF measurements which clearly required a two-component depolarization function in which both Gaussian and exponential depolarization terms were used to obtain accurate fits. The best fits to the time-dependent ZF depolarization data were given by

$$A(t) = A_1 \exp(-o^2 t^2) + A_2 \exp[-(\lambda t)^\beta] + A_{BG} \quad . (1)$$

The first term is a simple Gaussian depolarization function, where $\sigma$ is the width of the Gaussian frequency distribution (about zero). $\sigma$ is related to the mean square magnetic field variation at the muon site, $\left\langle \Delta h_\perp^2 \right\rangle_1$, via $\sigma = \gamma_\mu \sqrt{\left\langle \Delta h_\perp^2 \right\rangle_1}$ where $\gamma_\mu$ is the muon gyromagnetic ratio. The second term is a stretched exponential function, typically associated with depolarization due to rapidly fluctuating moments when $\beta \leq 1$. The third term results from muons implanted in the silver holder rather than the sample (30 – 35 % of the total initial asymmetry for the $Yb_4LiGe_4$ sample). Application of a 500 G LF completely suppresses the first term at low temperature ($\sigma \rightarrow 0$), while the second term is only moderately reduced, as expected for quasi-static and fluctuating contributions, respectively.

The $Yb_4LiGe_4$ data for temperatures 0.1 K $\leq T \leq$ 1.8 K were fit to Eq. 1, as shown in Fig. 7. At low temperature we determine that $A_1 \approx 0.9\ A_2$, and these values were found to be nearly constant over the temperature range cited above, and so were fixed. In Fig. 7 we plot the results for $\sigma$ and $\lambda$. For $T >$ 1.2 K $\sigma$ is nearly constant with a small value of 0.07 MHz, and is likely due to quasi-static Li nuclear moments. $\sigma$ increases abruptly below $T_{musr} =$ 1.2 K indicating the onset of depolarization by quasi-static magnetism, at a temperature below the peak temperatures in the AC susceptibility (Fig. 4) and specific heat (Fig. 5). The low temperature value of $\sigma =$ 0.3 MHz, corresponding to a root-mean-square field variation $\sqrt{\left\langle \Delta h_\perp^2 \right\rangle_1} = \dfrac{\sigma}{\gamma_\mu} \approx 3.5\ G$ .

The depolarization rate $\lambda$ increases monotonically with decreasing temperature, with no obvious structure near 1.2 K, attaining a value of 3.0 $\mu s^{-1}$ at $T =$ 0.1 K. The exponent $\beta$ begins increasing below 1 K from the higher temperature value of 0.7 to a value near 1 at low temperatures. Following Ref. 18, a quasi-static distribution would have a characteristic static local field of approximately $\lambda / \gamma_\mu \approx$ 35 G. Application of a longitudinal field of roughly five times this value, 165 G, would completely suppress quasi-static depolarization. However, we find that in an applied longitudinal field of 500 G the depolarization is still strong: at $T =$ 0.4 K, $\lambda =$ 0.9 $\mu s^{-1}$, compared to its value 2.8 $\mu s^{-1}$ in zero field, confirming that the origin of the second term in Eq. 1 is primarily dynamical in nature.

In Fig. 9 we show the temperature dependence of the parameter $\lambda$ extracted from data taken in 500 G, where depolarization due to quasi-static moments should be suppressed. We find a broad maximum near $T =$ 0.7 K. The maximum is characteristic of a fluctuating moment with a frequency that is decreasing with temperature and has a value on the order of the muon precessional frequency. Assuming that stretched exponential relaxation is due to field fluctuations in the strong collision limit with a single fluctuation rate $\nu$ [19],

$$\lambda = \frac{2\nu\gamma_\mu^2\left\langle\Delta h_\perp^2\right\rangle_2}{\nu^2 + \gamma_\mu^2 H^2} \quad , \quad (2)$$

where $H$ is the applied longitudinal field and $\left\langle\Delta h_\perp^2\right\rangle_2$ is the mean-square magnetic field variation for the muons experiencing the fluctuating field. At the maximum, $\nu = \gamma_\mu H = 43$ MHz, and using Eq. 2 we estimate $\sqrt{\left\langle\Delta h_\perp^2\right\rangle_2} \approx 80$ G. The temperature dependence of $\lambda$ is reasonably fit by Eq. 2 by using a power law variation of the frequency with temperature $\nu = AT^\alpha$, with $\alpha = 2.0(2)$ and $A = 94(5)$ MHz/K$^2$, as shown in Fig. 9. A small offset parameter $\lambda_0 = 0.07(2)$ $\mu$s$^{-1}$ was added to Eq. 2 in order to obtain a reasonable fit. Without a microscopic model, however, it is difficult to interpret this behavior.

There are several possible origins for the two component depolarization function. The muon could have two distinct stopping sites within a crystalline unit cell, but it is unlikely that the responses would be so dramatically different. Alternatively, for a magnetically ordered polycrystalline sample, in zero field on average 1/3 of the muons will have their moment lying along the field direction and will not precess [20]. However, this would constrain $A_1 = 2 A_2$, and is not consistent with our observed large value for $A_2$. Since neither of these possibilities seems likely, we infer that our sample is mesoscopically inhomogeneous; from our materials characterization we further conclude that the inhomogeneity is intrinsic.

We have not determined whether the inhomogeneity sets in at $T_{spht} = 1.75$ K, or if it persists to higher temperatures. Clearly our model implies the latter, since we found $A_1$ and $A_2$ to be nearly constant up to $T = 1.8$ K. However, at present we do not have a sufficiently detailed model to monitor the relative amplitudes of the two-component fit over the entire temperature range, and it is unclear if the apparent constant values for $A_1$ and $A_2$ are real, or artifacts of the approximate form of our depolarization function.

In Fig. 10 we show several depolarization curves for Yb$_5$Ge$_4$; the data only extend down to $T = 1.6$ K. Below 8 K we see a gradual increase in the depolarization rate. At $T = 8$ K the data show nearly exponential decay that is reduced, but not completely suppressed, by a 500 G applied longitudinal field. At $T$ = 2.1 K we begin to see a change in the magnetic state as evidenced by a quadratic curvature at short times. The data were fit to Eq. 1 (fits shown in Fig. 9) with $\beta \approx 1$, and we find that $\sigma$ and $\lambda$ increase monotonically below 8 K, reaching maximum values of 1.4 MHz and 0.43 $\mu s^{-1}$, respectively, at $T$ = 1.6 K. The amplitudes were related by $A_1$ = 0.25 $A_2$, i.e., the quasistatic component represented a significantly smaller fraction of the total response than for Yb$_4$LiGe$_4$. The monotonic behavior of both parameters is consistent with our AC-susceptibility measurements, which do not show any sign of ordering down to $T$ = 0.5 K. Following the analysis carried out for Yb$_4$LiGe$_4$, the reduction of $\lambda$ in a 500 G field yields a fluctuation frequency of 4.5 MHz at 1.6 K. A more definitive analysis of the behavior of Yb$_5$Ge$_4$ would require data taken to much lower temperatures.

## IV. DISCUSSION

Reviewing the important results, we find: the onset of partial magnetic ordering at $T_{spht}$ = 1.75 K; magnetic fluctuations that are evident at higher temperatures [10] and persist to temperatures $T \ll T_{spht}$ (a result consistent with the considerable entropy present at very low temperatures); and mesoscopic inhomogeneity associated with the magnetic response. The data for AC susceptibility, and $\mu$SR measurements also show abrupt changes below 2 K, but these occur significantly below $T_{spht}$ = 1.75 K, with $T_{spht} > T_{susc} > T_{musr}$. Specific heat is a semi-adiabatic (DC) measurement, while the susceptibility and $\mu$SR measurements probe frequencies in the kHz and MHz ranges, respectively, indicating that the fluctuations at these frequencies persist to lower temperatures. Similar behavior was observed by Olariu *et al*. [21] in the spin frustrated antiferromagnet NaCrO$_2$, who found changes in the $\mu$SR response at temperatures well below the Néel temperature as determined by specific heat measurements. This result was attributed to slowing of fluctuations in spin excitations persisting below the ordering temperature. Such a scenario seems likely in our case as well, albeit with a fairly rich fluctuation spectrum.

The fluctuation spectrum also includes components at frequencies beyond the $\mu$SR frequencies, as probed by NMR. In our earlier work [10], information regarding the nature of the high frequency fluctuations was obtained from the $T > 3$ K spin-lattice relaxation and specific heat data (see Figure 11). Following Moriya's self consistent renormalization theory (SCR) for quantum spin fluctuations [22], we observed a linear variation of $1/T_1T$ with $\chi^{3/2}$ below 30 K, which provided evidence for the presence of 2D ferromagnetic fluctuations in this range. The heat capacity data below 10 K provide further support for this assertion, as $C/T$ is observed to vary linearly with $T^{-1/3}$, also as predicted for 2D FM fluctuations. A similar analysis for the $1/T_1$ ($\lambda$) extracted from the $\mu$SR data does not follow the Moriya prediction, indicating that the lower frequency $\mu$SR is probing a different set of excitations; this gives further evidence for the existence of competing interactions in this system.

The FM correlations inferred from heat capacity and NMR data are surprising given the negative CW temperature extracted from high temperature magnetization and susceptibility data. The previously mentioned study by Ouyang *et. al* [7] on $Gd_5Ge_4$ has demonstrated that both FM and AFM correlations can coexist along different crystallographic directions, and lead to quite complex magnetic behavior. They observe a small static FM component at low temperatures inside the AFM matrix which grows with applied magnetic field, in addition to this they claim to observe the presence of dynamic FM clustering both above and below $T_N$ = 128 K which may be attributable to Griffiths-like state. In contrast, we do not observe multiple peaks in susceptibility or magnetization, and observe a strong suppression rather than an enhancement of the susceptibility peak with applied field. Nonetheless, their argument for the importance of interslab versus intraslab exchange interactions is quite appealing, as the addition of Li to the system introduces a magnetic-ion free layer inside the slab structure, possibly decreasing the effective dimensionality of the system. The facts that the Curie temperature, $\theta_{CW}$ for$Yb_4LiGe_4$ is 10 times smaller than for $Gd_5Ge_4$ and that $T_C$ is 100 times smaller suggest that quantum fluctuations become more important in the low temperature behavior of $Yb_4LiGe_4$.

The possibility of strong spin fluctuations is supported by both thermodynamic (heat capacity) and local ($1/T_1$) measurements, which agree well with SCR theory. We have calculated the SCR parameters $T_0$ and $T_A$ which characterize the width of the spin-fluctuation spectrum in frequency and $q$-space, respectively. We find ratios $T_A/T_0 \sim 160$, and $T_C/T_0 \sim 4$. Here $T_C$ is the Curie temperature associated with the high frequency fluctuations being probed by NMR, for which we have no reasonable estimate (it can be significantly different from the values inferred from $\mu$SR and susceptibility). Takahashi demonstrated that as both these ratios go to unity we recover the localized moment case [23]. Based on this, we see that $Yb_4LiGe_4$ could be described as an itinerant strongly correlated system.

Our results also show that $Yb_4LiGe_4$ shares many characteristics with other Yb intermetallic compounds that are characterized by a proximity to a magnetic instability. Following the scaling analysis of the magnetization data as discussed in Ref. 24, we have scaled our data to the form

$$M/H = T^{-\gamma} f(H/T^{\beta+\gamma}) \quad . \quad (3)$$

The results are shown in Fig. 3b, where we plot $MT^{\gamma}/H$ versus $H/T^{\beta+\gamma}$. Indeed, we find for $Yb_4LiGe_4$ between 2 K and 8 K that this form works quite well with parameter values of $\gamma = 0.6 \pm 0.1$ and $\beta = 0.3 \pm 0.1$. The magnetization data taken above 3 K for $Yb_5Ge_4$ also exhibit scaling with the similar exponents ($\gamma = 0.4 \pm 0.1$ and $\beta = 0.3 \pm 0.1$). This value for $\gamma$ is comparable to the value obtained for the quantum Griffiths phase material $CePd_{1-x}Rh_x$ [25], and matches the exponent determined from the static susceptibility in this region. However, our small value for $\beta$ (0.3 compared to 1.3 for $CePd_{1-x}Rh_x$) and the combined value of $\gamma + \beta \approx 1$ prevent us from distinguishing fluctuations originating from quantum critical behavior or local moment ordering [24].

While the results above originate from the competing interactions present in this system and the resultant magnetic frustration, the exact scenario by which this occurs is unclear. We will outline one possible scenario. Somewhere below 20 - 30 K phase separation is established, with some of the domains exhibiting FM correlations that eventually order at 1.75 K as seen in the specific heat. However, the other domains don't order and keep fluctuating. In these domains the fluctuations are slowing down and give rise to low $T$ peak in $\chi'$ and a rapid increase in $\sigma$, indicating the onset of quasi-static moments. The fact that the increase in $\sigma$ occurs at a lower temperature than that in $\chi'$ may indicate that the microscopic fluctuations are much faster than the macroscopic ones, namely that there are fast modes at $q > 0$. For example, if vortex-antivortex pairs are present [21] they will not contribute to the macroscopic susceptibility but will be detected by local probes such as muons.

## V. CONCLUSIONS

We have studied $Yb_4LiGe_4$, a compound related to the $R_5T_4$ family of materials whose magnetic properties have a high degree of sensitivity to the details of their layered crystal structure. Magnetic frustration produces a low temperature ground state characterized by mesoscopically inhomogeneous magnetic order setting in below $T = 1.75$ K and a complex spectrum of magnetic fluctuations that persist to very low temperatures. We also have made limited measurements on the parent compound $Yb_5Ge_4$. While we observe the broad peak near 3 K in AC susceptibility, which is a precursor to the reported magnetic ordering at 1.7 K observed through Mossbauer spectroscopy, we find no evidence for magnetic ordering in the AC susceptibility down to 0.5 K. These results may indicate that the low temperature ground state of this system is more complex than previously anticipated. Inelastic neutron scattering measurements on single crystals of $Yb_4LiGe_4$ and $Yb_5Ge_4$ would be particularly helpful in characterizing the nature of these fluctuations and the unusual low temperature ground states of both these materials.

**ACKNOWLEDGMENTS**

This work was supported by National Science Foundation Materials World Network grant DMR-0710525 (MJG) and REU grant DMR-0649169 (CB). Muon experiments were performed at the ISIS Muon Facility at the Rutherford Appleton Laboratories (UK) and the Swiss Muon Source at the Paul Scherrer Institute (Switzerland).

* Current address: Department of Physics, Brown University, Providence RI 02912

† Current address: Applied Physics Department, Rice University, Houston, TX 77005-1827

‡ grafm@bc.edu

**REFERENCES**

[1]     See for example Q. Si and F. Steglich, Science **329**, 1161 (2010).

[2]     For a review see "Introduction to Frustrated Magnetism," C. Lacroix, P. Mendels, and F. Mila, eds. (Springer-Verlag, Berlin, 2011).

[3]     M. S. Kim, M. C. Bennett, and M. C. Aronson, Phys. Rev. B **77**, 144425 (2008); M. S. Kim and M. C. Aronson, J. Phys.: Condens. Matter **23** 164204 (2011).

[4]     B. H. Bernhard, B. Coqblin, and C. Lacroix, Phys. Rev. B **83**, 214427 (2011).

[5]     V. K. Pecharsky and K. A. Gschneidner, Jr., Pure Appl. Chem., **79**, 1383 (2007).

[6]     B. S. Shastry and B. Sutherland, Physica B & C **108**, 1069 (1981).

[7]     Z. W. Ouyang, V. K. Pecharsky, K. A. Gschneidner, Jr., D. L. Schlagel, and T. A. Lograsso, Phys. Rev B **74**, 094404 (2006).

[8]     C. J. Voyer, D. H. Ryan, K. Ahn, K. A. Gschneidner, Jr., and V. K. Pecharsky, Phys. Rev. B **73**, 174422 (2006).

[9]     Q. Xie, C. Kubata, M. Worle, and R. Nesper, Z. Anorg. Allg. Chem., **634**, 2469 (2008).

[10]    S. C. Peter, S. M. Disseler, J. N. Svensson, P. Carretta, and M. J. Graf (submitted for publication; arXiv:1109.6144).

[11]    K. Ahn, A. O. Tsokol, Y. Mozharivskyj, K. A. Gschneidner, Jr., V. K. Pecharsky, Phys. Rev. B **72**, 054404 (2005).

[12]    P. Tobash and S. Bobev, J. Am. Chem. Soc. **128** 3525 (2006).

[13]    P. Rhodes and E. P. Wohlfarth, Proc. R. Soc. London, Ser. A **273**, 247 (1963).

[14]    R. M. Moon, W. C. Koehler, H. R. Child, and L. J. Raubenheimer, Phys. Rev. 176, 722 (1968).

[15]    A. Amato, Rev. Mod. Phys **69,** 1119 (1997).

[16]    S. Blundell, Contemporary Physics **40,** 175 (1999).

[17]    *Muon Spin Rotation, Relaxation, and Resonance : Applications to Condensed Matter*, A. Yaouanc and P. Dalmas de Réotier, (Oxford University Press, Oxford, 2011).

[18]    F. Bert, P. Mendels, A. Olariu, N. Blanchard, G. Collin, A. Amato, C. Baines, and A. D. Hillier, Phys. Rev. Lett. **97**, 117203 (2006).

[19]    R. S. Hayano, Y. J. Uemura, J. Imazato, N. Nishida, T. Yamazaki, and R. Kubo, Phys. Rev. B **20**, 850 (1979).

[20]     For a recent discussion, see F. L. Pratt, J. Phys.: Condens. Matter **19**, 456207 (2007) and references therein.

[21]     A. Olariu, P. Mendels, F. Bert, B. G. Ueland, P. Schiffer, R. F. Berger, and R. J. Cava, Phys. Rev. Lett. **97**, 167203 (2006).

[22]     A. Ishigaki and T. Moriya, J. Phys. Soc. Jpn. **67**, 3924 (1998).

[23]     Y. Takahashi, J. Phys. Soc. Jpn. **55**, 3553 (1986).

[24]     A. M. Tsvelik and M. Reizer, Phys. Rev. B **48**, 9887 (1993).

[25]     D. T. Adroja, A. D. Hillier, J.-G. Park, W. Kockelmann, K. A. McEwen, B. D. Rainford, Kwang-Hyun Jang, C. Geibel, and T. Takabatake, Phys Rev. B **78**, 014412 (2008).

FIGURE CAPTIONS

Fig 1. Projections of the crystal structure of $Yb_4LiGe_4$ viewed along (a) the a-axis to highlight the effect of Li substitution and (b) the b-axis to show Yb atoms on a Shastry-Sutherland lattice.

Fig. 2. Temperature dependent resistance of $Yb_4LiGe_4$ below 2 K and in applied fields up to 4 T. Inset shows resistivity up to 300 K in zero magnetic field (adapted from Ref. 10).

Fig. 3 (a) Magnetization isotherms of $Yb_4LiGe_4$ between 2 K and 10 K. (b) Scaling of magnetization isotherms for $Yb_4LiGe_4$ following the procedure outlined in the text.

Fig. 4. Real part of ac-susceptibility below 2 K for $Yb_4LiGe_4$. Insert shows ac susceptibility for $Yb_5Ge_4$ above 0.5 K for comparison.

Fig. 5. Specific heat below 3 K in zero and applied magnetic fields.

Fig. 6. Depolarization curves for $Yb_4LiGe_4$ in zero magnetic field and 500 G at several temperatures. Solid lines are fits to Eq. (1) as described in the text.

Fig. 7. Evolution of depolarization below 1.8 K for $Yb_4LiGe_4$. Data for 0.4 K and 1.8 K are shown, while lines are fits using Eq. 1 to data taken at 200 mK intervals. Inset shows the short-time data from PSI at a base temperature of 20 mK.

Fig. 8. Temperature dependence of the depolarization rate $\lambda$, and frequency distribution width $\sigma$, for $Yb_4LiGe_4$. Inset: The temperature variation of the exponent $\beta$, as defined in Eq. 1.

Fig. 9. Temperature dependence of the depolarization rate $\lambda$ in an applied longitudinal field of 500 G for $Yb_4LiGe_4$. The solid line is a fit to Eq. 2 using a quadratic variation of the fluctuation frequency with temperature, as described in the text.

Fig. 10. Depolarization curves for $Yb_5Ge_4$ at several temperatures in ZF and 500 G.

Fig. 11. (a) Temperature variation at intermediate temperatures of the specific heat and (b) spin-lattice relaxation rate variation with DC susceptibility, for $Yb_4LiGe_4$. Both curves are consistent with two-dimensional ferromagnetic fluctuations, as described in the text (adapted from Ref. 10).

Figure 1

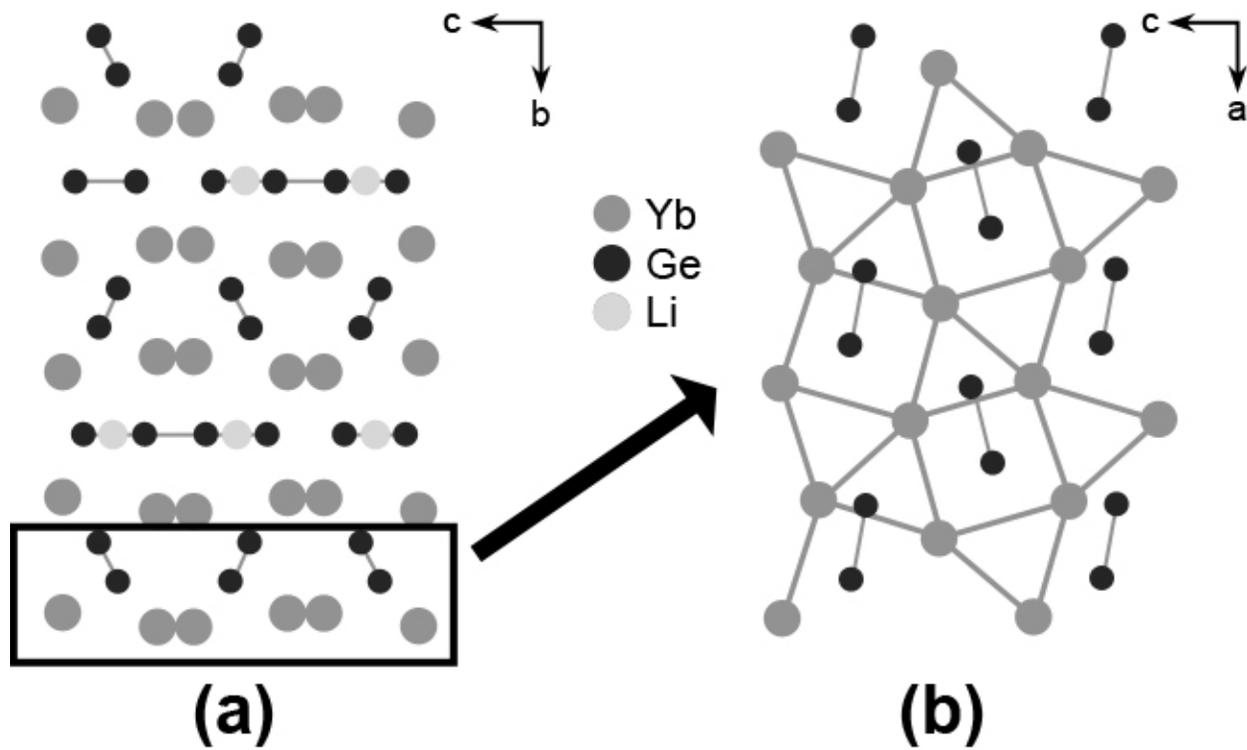

**(a)**                    **(b)**

Figure 2

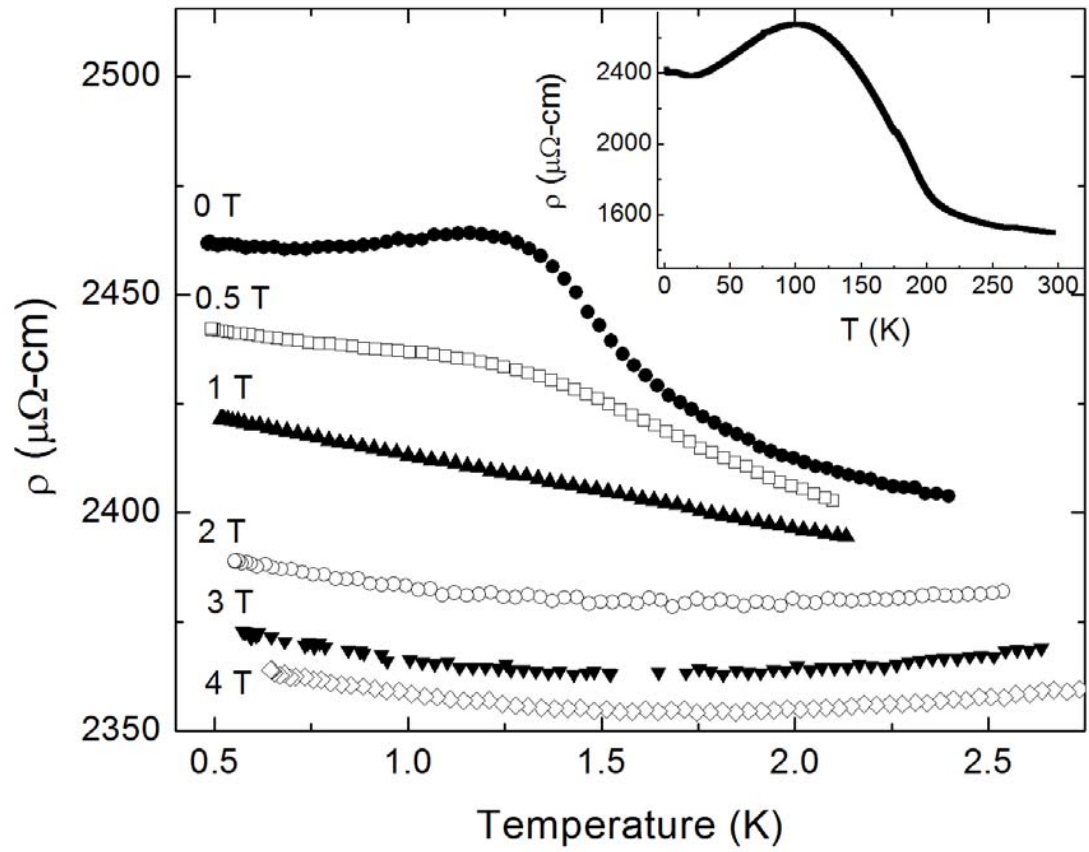

Figure 3

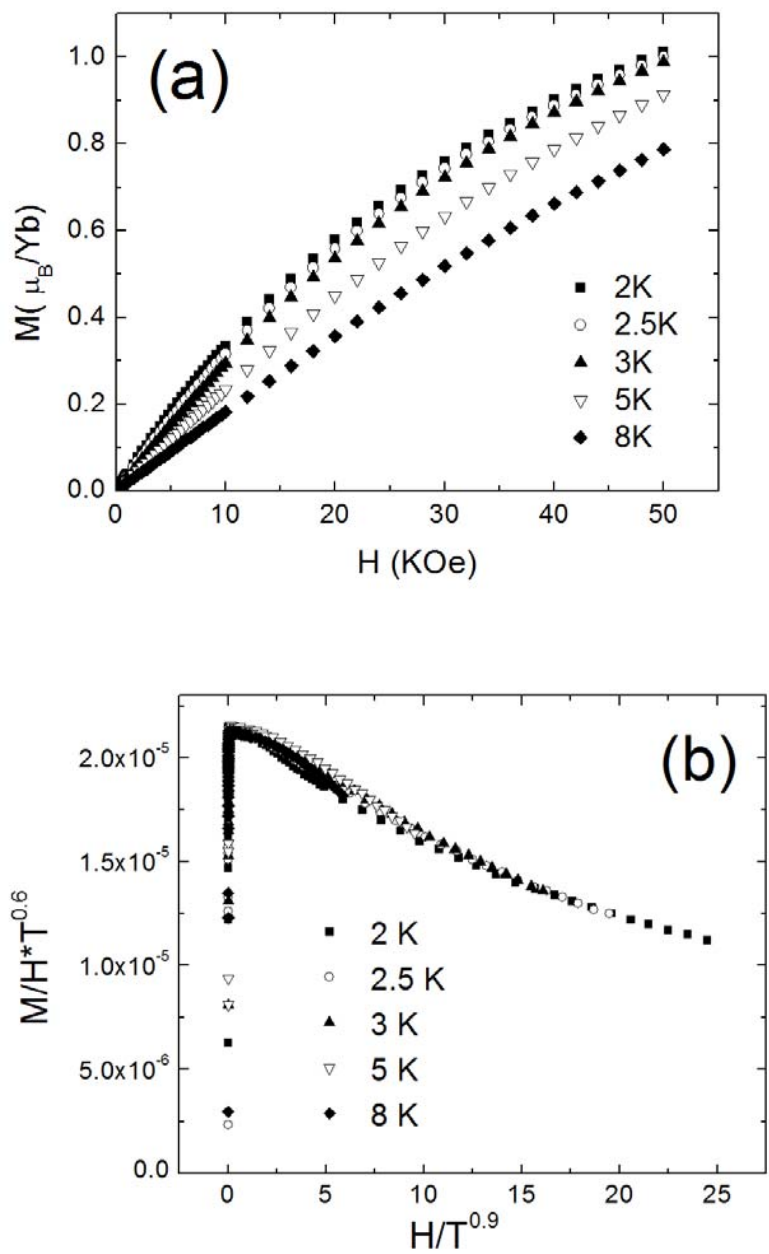

Figure 4

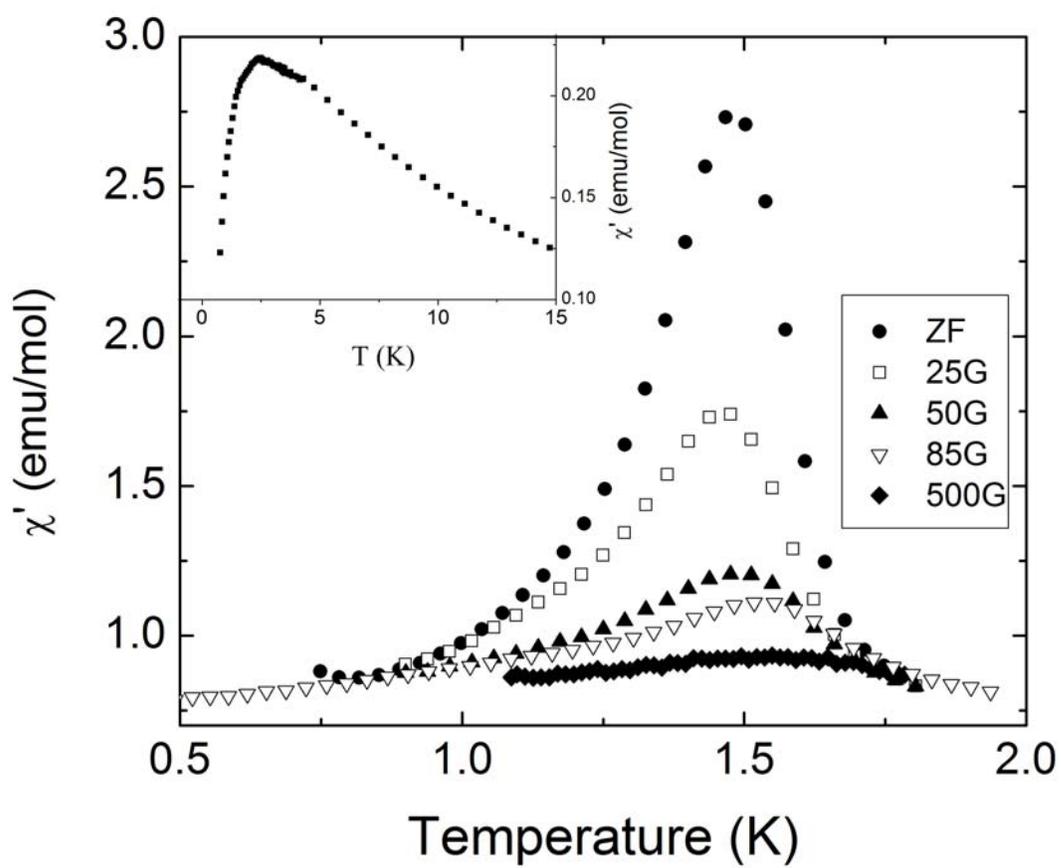

Figure 5

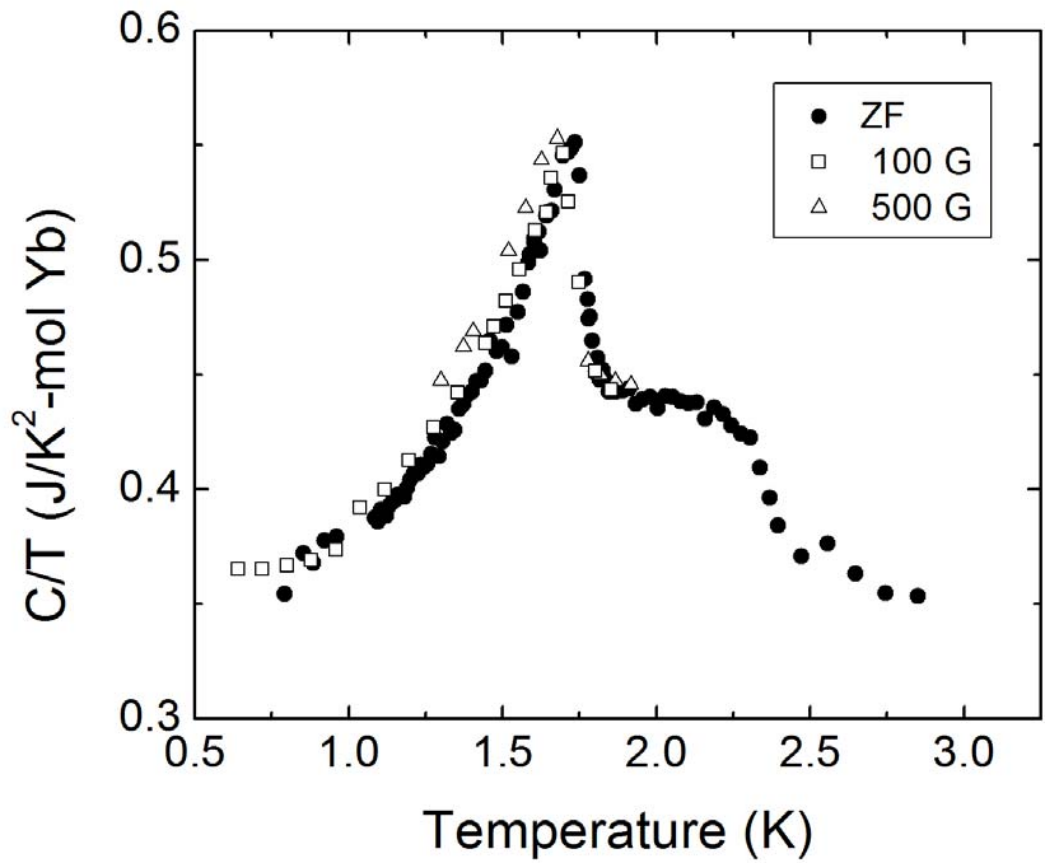

Figure 6

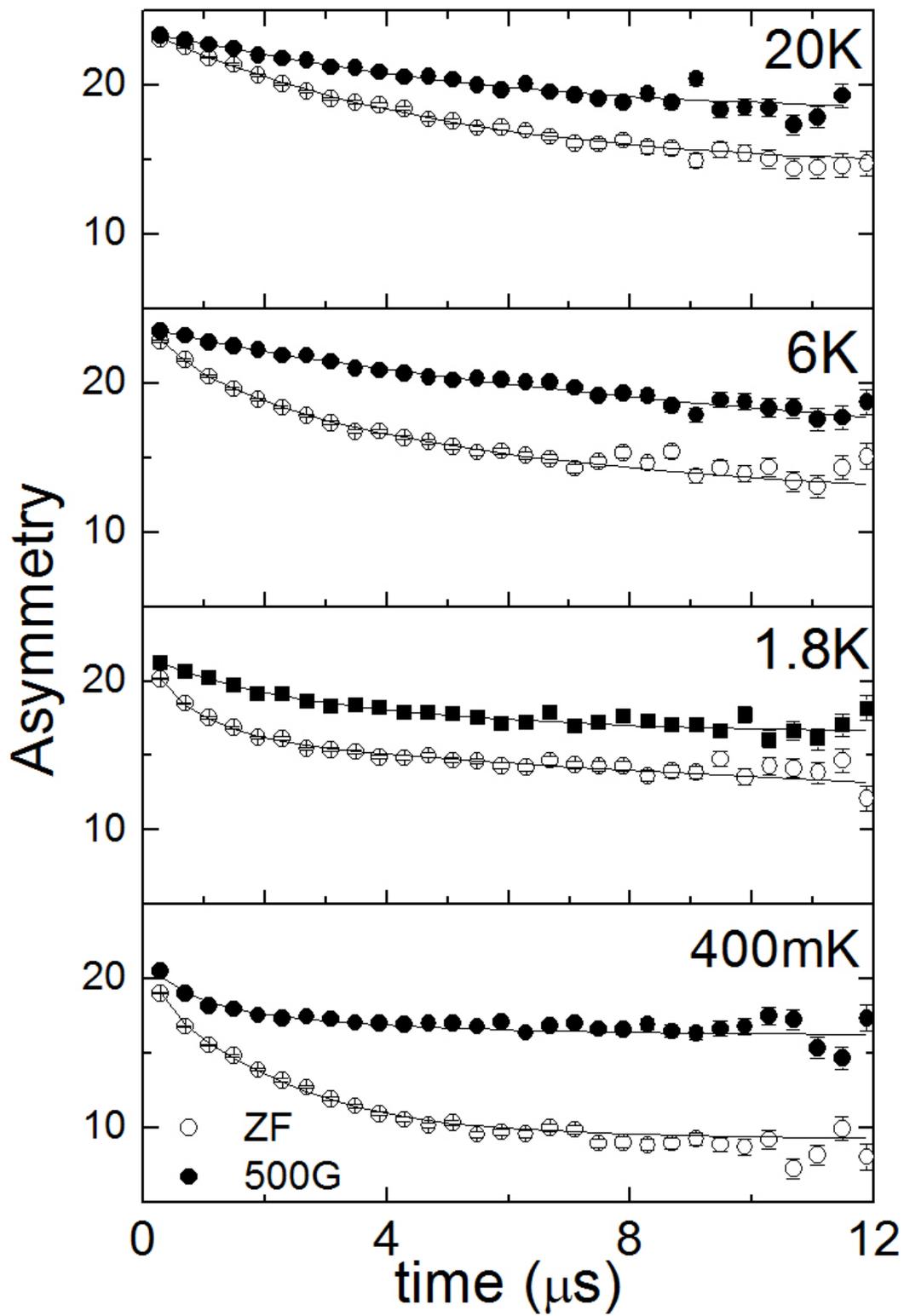

Figure 7

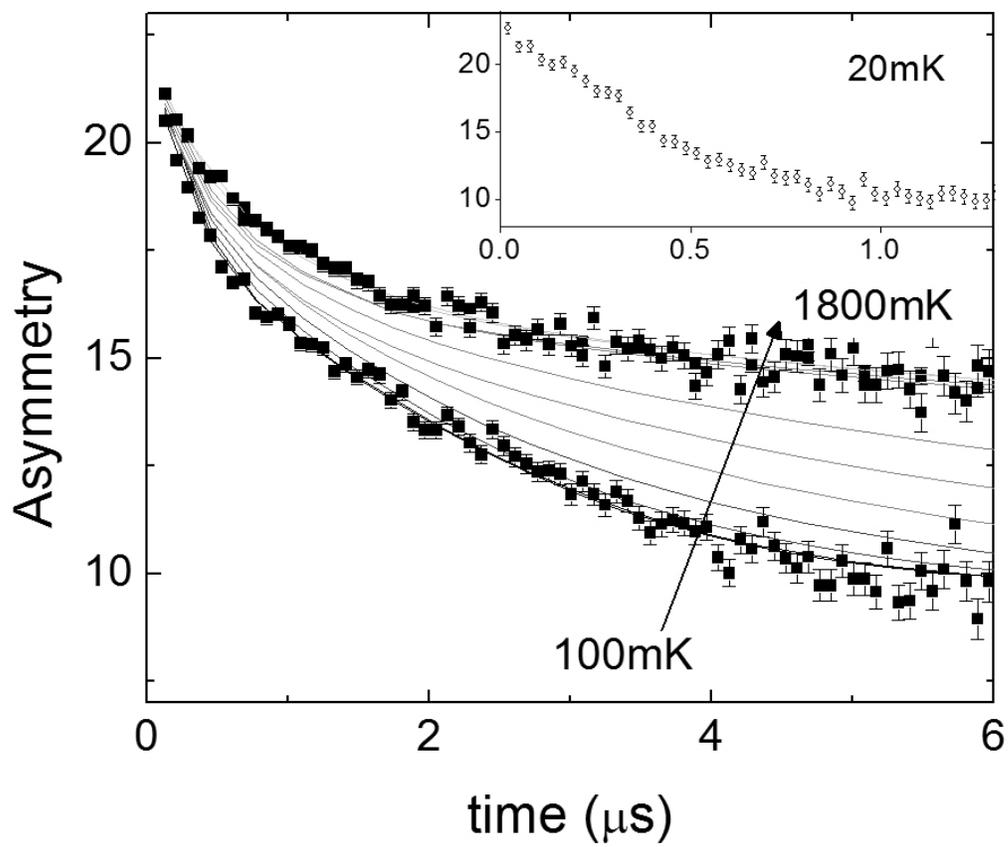

Figure 8

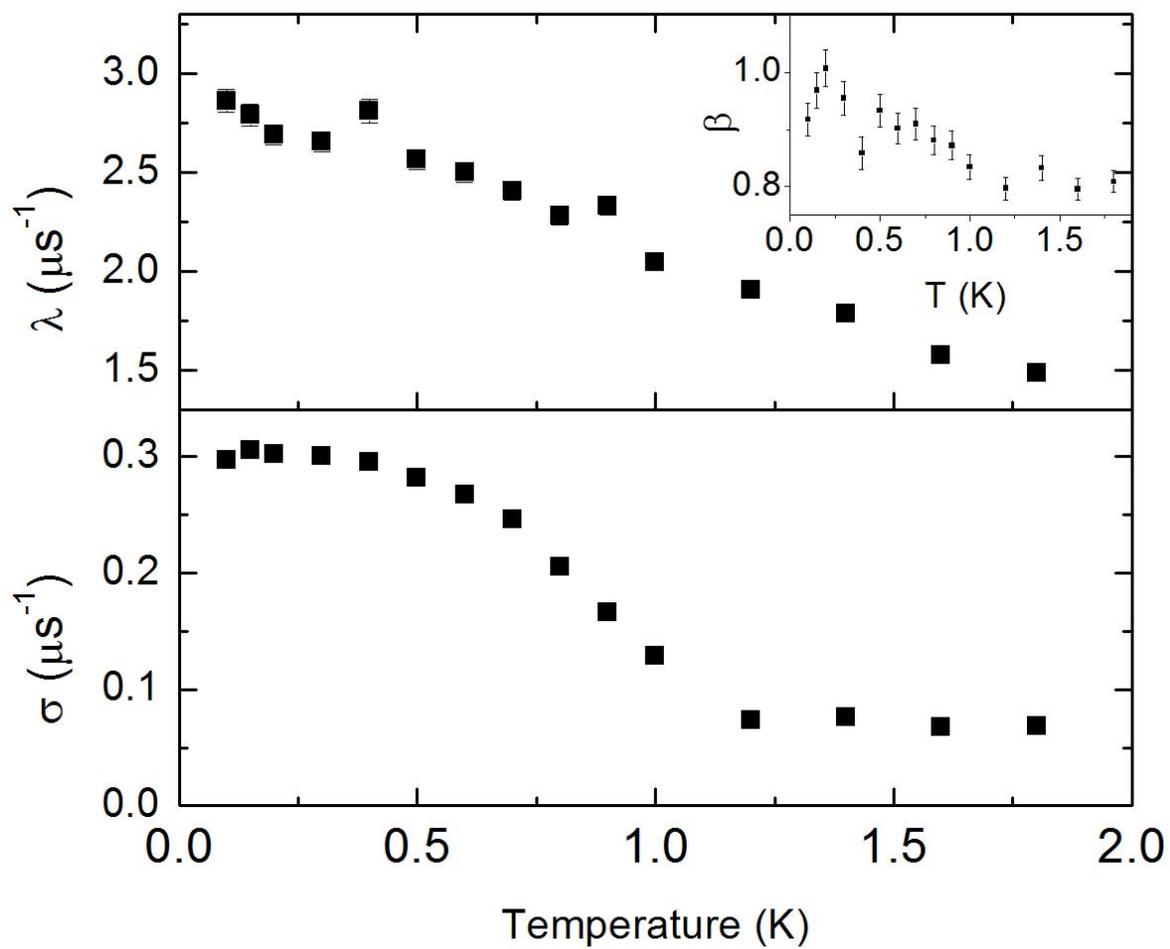

Figure 9

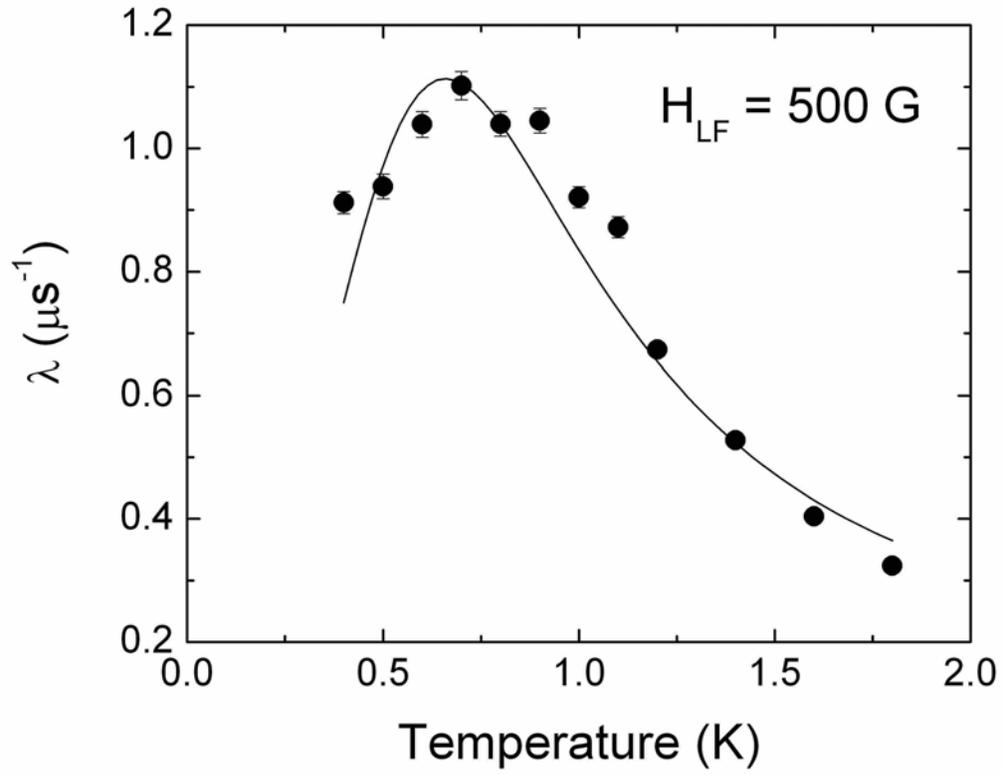

Figure 10

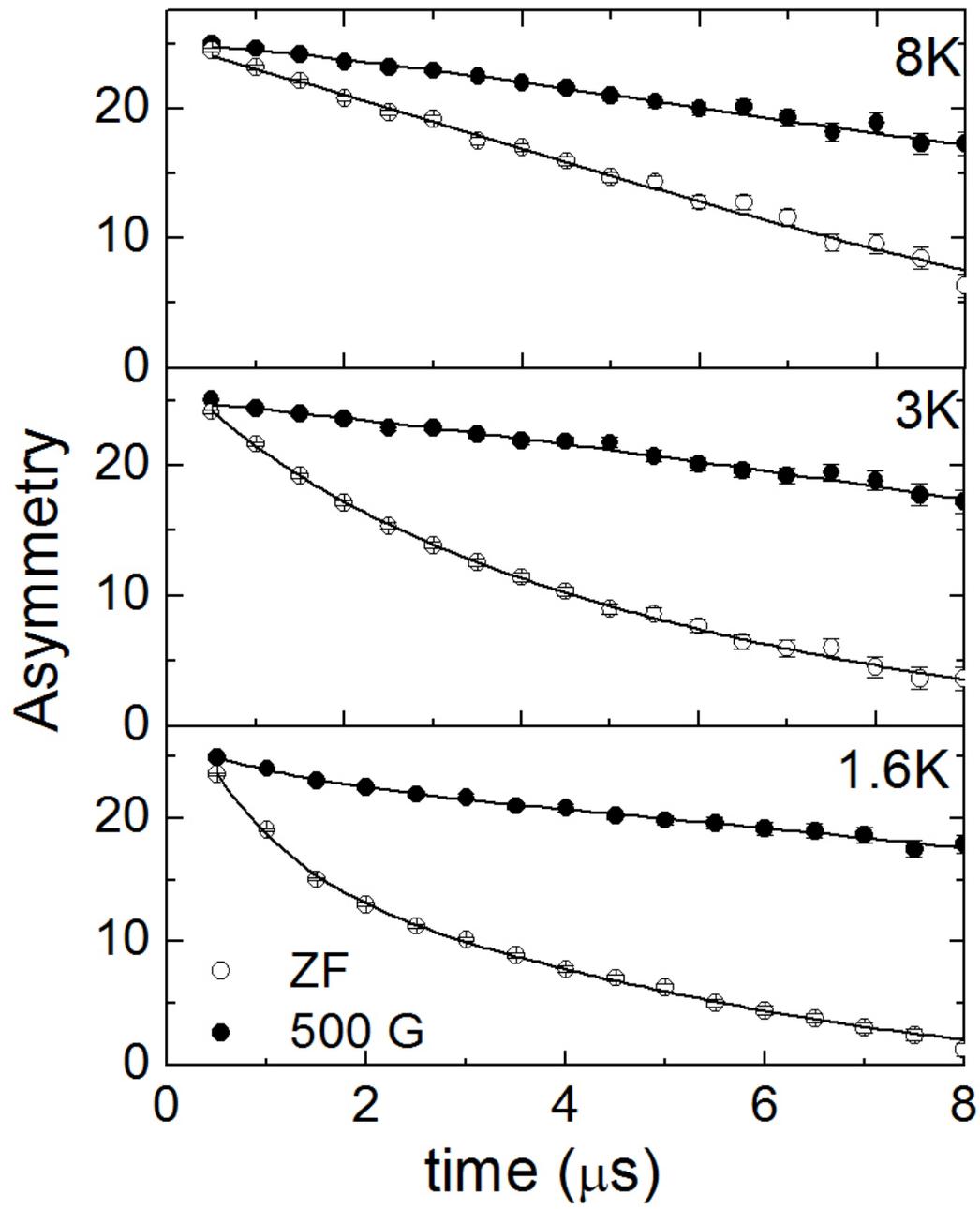

Figure 11

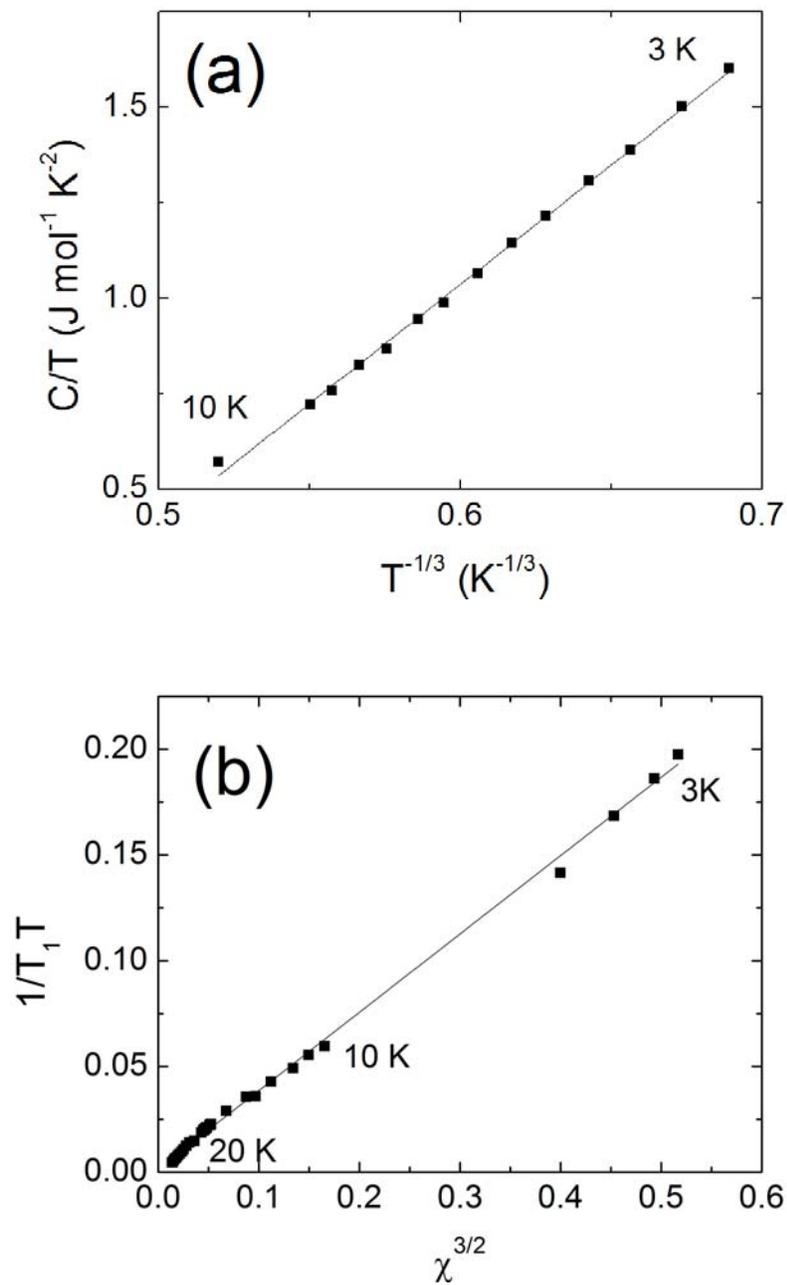